\documentclass[sigconf]{acmart}

\usepackage{multirow}

\AtBeginDocument{%
  }

\copyrightyear{2024} 
\acmYear{2024} 
\setcopyright{rightsretained} 
\acmConference[FDG 2024]{Proceedings of the 19th International Conference on the Foundations of Digital Games}{May 21--24, 2024}{Worcester, MA, USA}
\acmBooktitle{Proceedings of the 19th International Conference on the Foundations of Digital Games (FDG 2024), May 21--24, 2024, Worcester, MA, USA}
\acmDOI{10.1145/3649921.3649992}
\acmISBN{979-8-4007-0955-5/24/05}

\begin{document}
%%
%% The "title" command has an optional parameter,
%% allowing the author to define a "short title" to be used in page headers.
\title{GameDevDojo - An Educational Game for Teaching Game Development Concepts}

%%
%% The "author" command and its associated commands are used to define
%% the authors and their affiliations.
%% Of note is the shared affiliation of the first two authors, and the
%% "authornote" and "authornotemark" commands
%% used to denote shared contribution to the research.
\author{Michael Holly}
\affiliation{%
  \institution{Graz University of Technology}
  \city{Graz}
  \country{Austria}
}
\email{michael.holly@tugraz.at}

\author{Lisa Habich	}
\affiliation{%
  \institution{Graz University of Technolog}
  \city{Graz}
  \country{Austria}
}
\email{lisa.habich@tugraz.at}

\author{Johanna Pirker}
\affiliation{%
  \institution{Ludwig-Maximilians-Universität München}
  \city{Munich}
  \country{Germany}
}
\email{jpirker@iicm.edu}

%%
%% By default, the full list of authors will be used in the page
%% headers. Often, this list is too long, and will overlap
%% other information printed in the page headers. This command allows
%% the author to define a more concise list
%% of authors' names for this purpose.
\renewcommand{\shortauthors}{Holly, et al}

%%
%% The abstract is a short summary of the work to be presented in the
%% article.
\begin{abstract}
Computer Science (CS) has experienced significant growth and diversification in recent decades. However, there is a lack of diversity in CS learning approaches. Traditional teaching methods and hands-on learning dominate this field, with limited use of playful and interactive learning methods such as educational games. This gap is particularly evident in game development as a subfield of CS. To address this problem, we present a game-based learning approach to teach foundational concepts for game development. The paper aims to expand the educational landscape within CSE, offering a unique and engaging platform for learners to explore the intricacies of game creation by integrating gamified learning strategies. In this paper, we investigate the user's learning experience and motivation, and the differences between traditional learning and game-based learning methods for teaching game development concepts. The study involves 57 participants in an AB test to assess learners' motivation, user experience, and learning outcomes. The results indicate a significantly increased learning outcome for the game-based learning approach, as well as higher motivation in learning game development concepts.
\end{abstract}

%%
%% The code below is generated by the tool at http://dl.acm.org/ccs.cfm.
%% Please copy and paste the code instead of the example below.
%%


%%
%% Keywords. The author(s) should pick words that accurately describe
%% the work being presented. Separate the keywords with commas.

\received{30 November 2023}
%\received[revised]{12 March 2009}
\received[accepted]{23 February 2009}

%%
%% This command processes the author and affiliation and title
%% information and builds the first part of the formatted document.
\maketitle

\section{Introduction}

Innovative methods are constantly sought to effectively teach complex concepts and skills in the rapidly evolving field of computer science education (CSE). Various strategies and approaches have been formulated to enhance educational practices within computer science (CS). Papastergiou \cite{papastergiou2009digital} demonstrated the efficacy of digital gaming concepts in boosting motivation while learning computer memory principles. Incorporating interactive simulations, visual aids, and gamified learning environments is a promising approach to improve teaching methods in this field \cite{garneli2018programming}. Especially in the context of computer science, simulation environments have proven to be effective tools for simplifying complex ideas and making them more tangible and accessible \cite{alnoukari2013simulation, wolff2000satsim}. This variety of innovative approaches underscores the dynamic evolution of educational paradigms that aim to meet the ever-changing preferences and requirements of learners. Teaching game development involves various CS aspects, ranging from programming and algorithms to graphics rendering and user interaction \cite{gregory2018game}. It has evolved into an engaging and pragmatic method that not only captures students' attention, but also provides a tangible framework for understanding programming, problem-solving, and design principles. Taking advantage of the features of games, such as interactivity, challenge, and reward, game-based learning promotes active engagement and experiential understanding. This approach is especially important in technical fields such as programming and game development, where hands-on experience is often essential. Integrating game elements into educational contexts not only enriches the learning process, but also enhances learners' motivation, critical thinking, and problem-solving skills \cite{gurbuz2022serious}.
Games with a purpose use the inherent appeal of games to create immersive learning experiences. They often incorporate elements like challenges, rewards, competition, and simulation to make learning engaging and enjoyable. This approach can be particularly effective in capturing the attention of learners, fostering active participation, and improving retention of the material being taught. There are already many tools like Scratch, Snap!, Alice, or Greenfoot that use exciting gamified activities to introduce programming to K-12 students. Miljanovic and Bradbury \cite{miljanovic2018} already showed the gap of educational games in non-fundamental CS concepts like development methods or software design. In particular, there is a lack of games to teach the development of games themselves in a playful way without programming.

In this paper, we introduce an innovative and immersive educational game designed to equip aspiring game developers with a comprehensive understanding of game development concepts. The users dive into a virtual world where they encounter challenges that require them to apply different game development concepts. Each achievement unlocks more intricate aspects of game development, guiding the user toward mastery. 
To evaluate this learning approach, we defined the following research objectives:
\begin{itemize}
    \item Investigating the user learning experience and motivation of a game-based learning tool to teach game development concepts. 
    \item Analyzing the differences between traditional learning and a game-based learning approach to teach fundamentals in game development.
\end{itemize}

\paragraph*{Contribution.}
This paper introduces an AB study with 57 middle and high school students evaluating a game-based learning tool to teach fundamental concepts of game development. The study aims to provide information on the effectiveness of the game-based learning approach used by the GameDevDojo tool compared to a traditional classroom-based approach with a focus on the learning experience and motivation.
\section{Background and Related Work}

Many people often associate computer science with programming or coding and miss the broad scope of CS. The increasing demand for computer science makes it necessary to engage people in this field and to find interesting and motivating learning methods. Nowadays, educators often use theoretical lecture-based teaching methods and project-based strategies. However, to achieve a comprehensive computer science education, different teaching approaches are needed. The learning approach by Bell \cite{Bell1998} uses unplugged activities to teach different CS concepts without a computer. Such activities can include games or cards intended to demonstrate computer science concepts and support students in their conceptual understanding. Although unplugged learning aims to understand computer science concepts without relying on technology by incorporating hands-on and tangible activities, interactive digital experiences designed for a purpose promote skill development and conceptual understanding through immersive learning. The visual programming environment Scratch allows users to learn programming while creating their own interactive stories, games, and animations using a visual programming language \cite{maloney2010scratch}. As an extended implementation of Scratch, the tool Snap! allows the creation of custom blocks and control structures. These features add the capabilities to make it more suitable for serious introductions to computer science \cite{RomagosaiCarrasquer2019}. The block-based programming environment Alice further enables the easy creation of animations, interactive narratives, and the development of simple 3D games through creative exploration. This approach has been shown to have a positive impact on performance and retention \cite{cooper2000alice, cooper2003using}. Another tool specialized in the development of interactive graphical applications is Greenfoot. It enables high school students to develop engaging and interesting programs in a fast and easy way while learning fundamental programming concepts \cite{kolling2010greenfoot}. All of these environments can be collectively categorized due to their shared characteristics. They are all visual and designed to create immediate engagement through exciting activities and focus on introducing programming to K-12 students.
The potential of video games for educational purposes has been proven over decades and can be used to teach different subjects. Thereby, the right balance has to be achieved between the knowledge that is taught and the gameplay \cite{plass2015foundations}. Especially for serious games which refer to interactive digital applications that go beyond mere entertainment and serve specific social, educational, or problem-solving purposes. This makes it only natural to use such games as tools for game-based learning \cite{Schrader2022}. Lameras et al. \cite{lameras2017essential} establish a taxonomy connecting learning and game attributes, offering insights for practitioners in designing serious games. The taxonomy guides serious game and instructional designers, game developers, academics, and students on representing learning elements such as activities, outcomes, feedback, and assessment within games.
While the use of achievement systems is a common practice in entertainment games to provide feedback and rewards, it has been shown that embedding achievements in serious games has the potential to increase engagement and game time \cite{cannon2009synthetic}. By providing rewards for completing tasks, achievements can effectively encourage learners to invest more time in the game. Research has shown that dedicating increased effort and time to a task can amplify the educational value derived from the experience \cite{fisher1998differential}. Additionally, the establishment of achievement-oriented goals plays a pivotal role in motivating individuals within contexts focused on accomplishment \cite{elliot2005conceptual}.

Research in the field of computer science has shown that there are multiple kinds of CS games with a focus on teaching programming \cite{Vahldick2014, miljanovic2018}. 
Furthermore, Papastergiou \cite{papastergiou2009digital} showed different digital gaming concepts to enhance motivation when learning about computer memory principles. Another game inspired by ESA's ExoMars mission is the educational game NEPO\footnote{https://www.roberta-home.de/en/}, where the goal is to detect signs of life on Mars while programming and controlling a simulated rover. Miljanovic and Bradbury \cite{miljanovic2018} evaluated such serious game approaches concerning likability, accessibility, learning effect, and engagement. Their results show that many of the analyzed games focus primarily on problem-solving and fundamental programming concepts. There is a noticeable absence of games that address topics such as data structures, development methods, and software design. Furthermore, there is a lack of serious games with a focus on game development. One way to engage students in computer science education is through game development using a game engine. Comber et al. \cite{comber2019engaging} utilized the Unity game engine to instruct secondary school students in the creation of their own video games. Their results suggest that Unity may not be the optimal choice as the primary tool for introductory instruction. For students, it is difficult to deal with the vast amount of options provided by the engine and to dive into more complex coding structures and patterns. Unity Bolt (visual scripting) attempts to address these issues by allowing the creation and development of gameplay mechanics using a visual, graph-based system instead of writing code \cite{unitybolt}. The use of visual scripting tools can increase motivation and understanding of iteration and sequencing as fundamental concepts in computer science \cite{Duncan2019, duncan2015pilot}. Furthermore, tools like the OniStep\footnote{https://onistep.com/} plugin can be used to support beginners to facilitate the seamless creation of new projects in a game development environment. It utilizes artificial intelligence technologies to create a step-by-step tutorial so that users are no longer burdened with navigating complicated settings or extensive documentation. Engaging students in the process of creating games can increase their interest in the subject, foster a sense of ownership, and lead to a deeper appreciation of the knowledge they have learned. Be{\c{c}}a et al. \cite{becca2020promoting} developed a toolkit to engage students in the design of digital games. It includes resources and tools to encourage active participation in game-creation activities. Initial findings suggest that the toolkit could be considered an engaging method for game creation and could be implemented in various education topics.

\section{Learning Application}

\begin{figure}[t]
    \centerline{\includegraphics[width=0.4\textwidth]{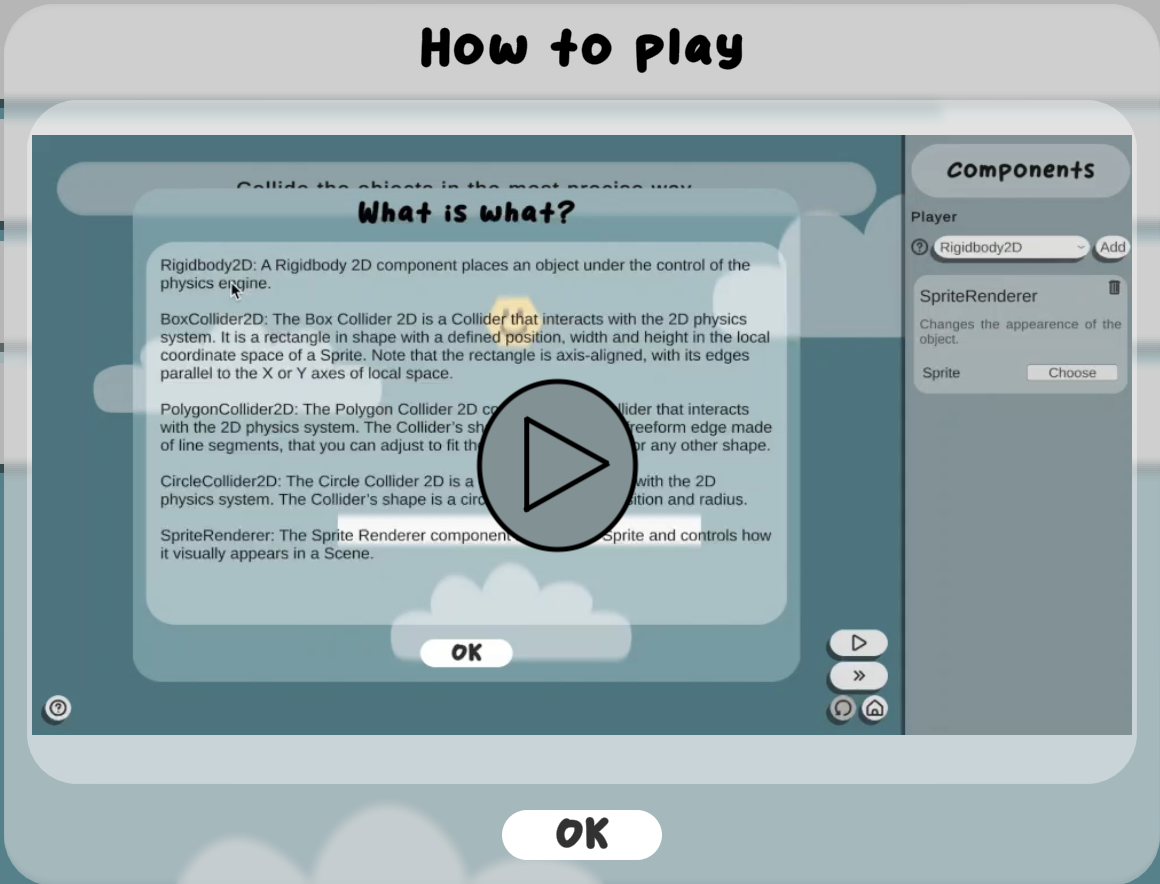}}
    \caption{The Manual View, providing a video introducing the interaction with the application.}
    \Description{The Manual View, providing a video introducing the interaction with the application.}
    \label{fig:howto}
\end{figure}

The learning application is developed in Unity\footnote{https://unity.com/} and is based on its physics engine allowing the user to manipulate game objects during the gameplay. The app is designed as an interactive and engaging learning platform to teach users game development concepts by engaging them in practical scenarios and fostering a deep understanding of the fundamental principles of game development. It provides an intuitive interface, enabling users to interact with predefined game objects, adjust settings, and witness real-time effects, thereby creating a dynamic and effective learning experience.
%1.2
The game does not have any specific prerequisites for the players and is self-explanatory. It provides a manual in the main menu which shows a short video on how to use and interact with the game and the different components. Figure \ref{fig:howto} illustrates the manual view showing a video that introduces the interactions with the application.

\subsection{Game Design and Mechanics}
\begin{figure}[t]
    \centerline{\includegraphics[width=0.5\textwidth]{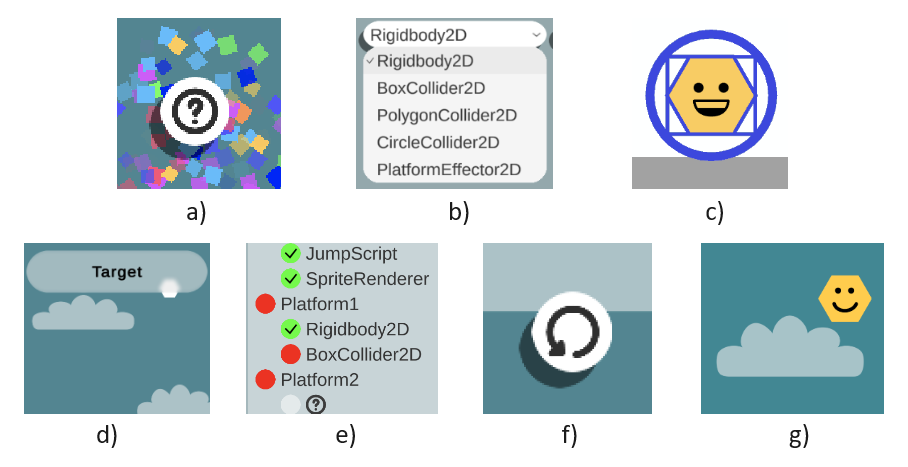}}
    \caption{Gamification elements: a) Acknowledgement, b) Imposed Choice, c) Novelty, d) Objectives, e) Progression, f) Renovation, g) Sensation}
    \Description{Gamification elements: a) Acknowledgement, b) Imposed Choice, c) Novelty, d) Objectives, e) Progression, f) Renovation, g) Sensation}
    \label{fig:game_elements}
\end{figure}

%1.1
To enhance the understanding of development principles, we implemented selected concepts based on the gamification taxonomy of Toda et al.~\cite{toda2019taxonomy}, along with training and education methods and strategies introduced by Kapp~\cite{kapp2012gamification}. The learning application incorporates seven key elements: (1) \textit{Acknowledgement}, (2) \textit{Imposed Choice}, (3) \textit{Novelty}, (4) \textit{Objectives}, (5) \textit{Progression}, (6) \textit{Renovation}, and (7) \textit{Sensation}. The acknowledgment feature in the game is represented by confetti raining around the help menu button when a correct configuration of a component is achieved, the imposed choice aspect is related to selecting the components, and the collider visualization embodies the novelty element, allowing immediate changes to components throughout the level. The level targets represent the objective element and progression is conveyed through the help menu, updating players on component statuses. Furthermore, renovation is facilitated by restarting within the level or replaying unlocked levels, and sensation is enhanced by creating a comic-like atmosphere with sky scenes and customizable facial expressions for players. The implemented elements and their representations are shown in Figure \ref{fig:game_elements}.

\subsection{Level Setup}

\begin{figure}[t]
    \centerline{\includegraphics[width=0.5\textwidth]{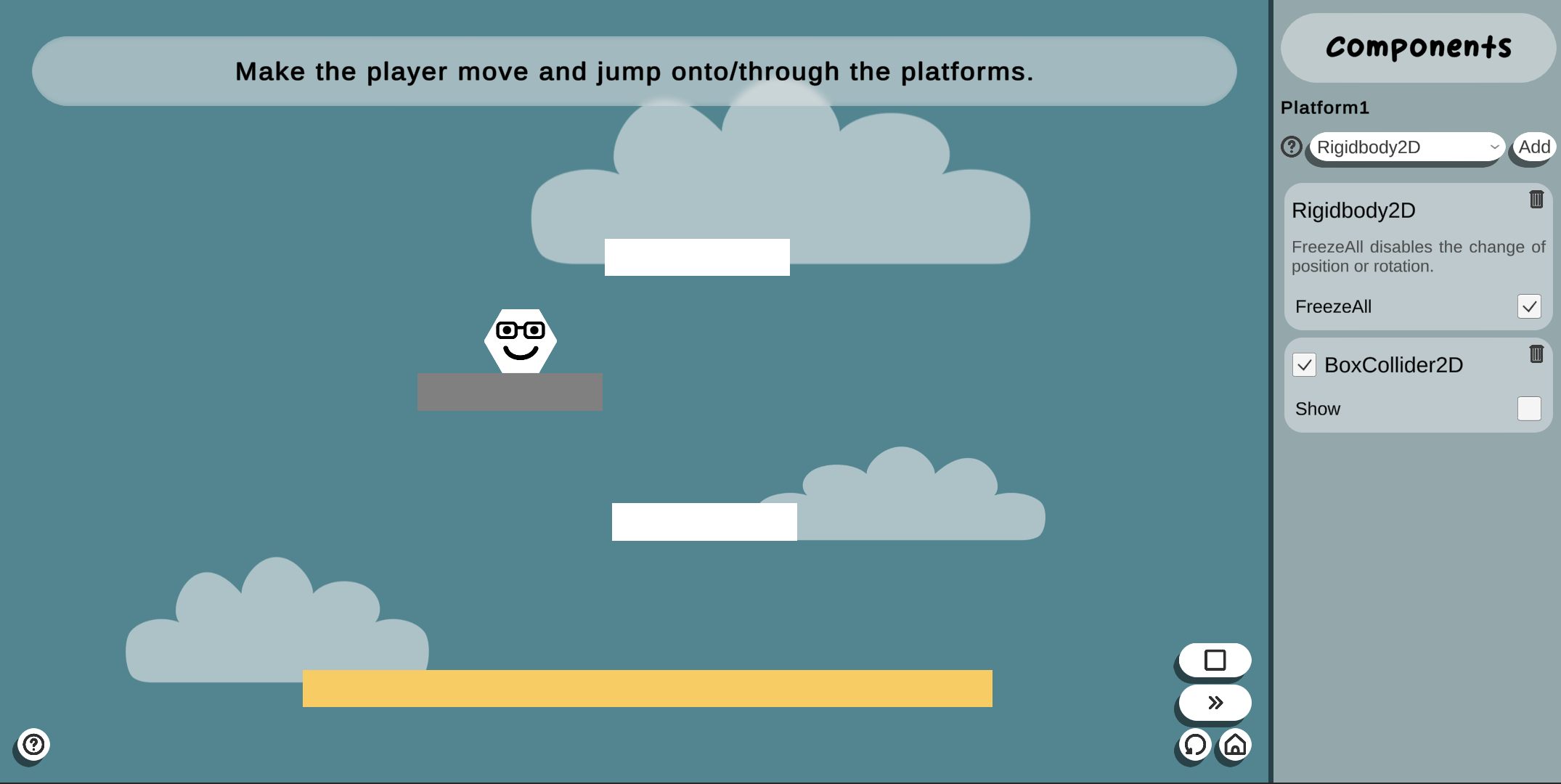}}
    \caption{GameDevDojo-Level Design, showing the game scenario, the goal of the level, the component view, and control buttons.}
    \Description{GameDevDojo-Level Design, showing the game scenario, the goal of the level, the component view, and control buttons.}
    \label{fig:level_design}
\end{figure}

The learning application consists of a level selection menu and different learning activities in the form of game levels. The main menu allows the player to choose unlocked levels that become accessible after completing the task. The levels are designed for repetitive learning and experimentation. Each level has a clear goal description displayed at the top of the screen, a modification view to add and change game components on the right side, and control buttons placed in the corner to start, stop, and validate the solution (see Figure \ref{fig:level_design}). The application includes five progressive game levels designed to be both repetitive and challenging. 
The initial level focuses on changing a game object's sprite and adding gravity. The second and third level include different kinds of collisions, whereas the latter one also brings attention to memory usage. In Level 4 the player has to add player movement with pre-defined scripts, as well as platform effectors to jump onto/through game objects. The last level deals with triggers and UI displays.

\subsection{Gameplay}

To reach the goal of the level, the player has to manipulate the predefined objects in the level by either adding new game components or adjusting the component settings. 
%1.3
The components the user must add are shown in a separate view and a clear description of the components can be accessed via the (?)-button next to the component. Adding components can be accomplished by clicking on a game object, which opens the corresponding settings in the component view. To attach a component, the user can choose between a variety of options accessible via a dropdown menu. By pressing the "add"-button the selected component is then added to the desired object. In case the user is unsure about the appropriate choice or the function of a component, additional information can be acquired via the question mark next to the component selection. To modify the settings, the component can be selected or unselected by clicking on the corresponding entry in the component list. By toggling the check box next to the component name, users can enable and disable specific elements. Therefore, deactivating a component has the same result as removing the component from the list. To support the player in reaching the learning objectives, specific components provide concise explanations of their corresponding values. If the player configures a component correctly, a small confetti rain appears around the "help"-button in the left corner to indicate that the user has completed a subtask and achieved a part of the level outcome. To give a better insight into the in-game component(s)/-groups during the gameplay, the most important ones are outlined below.

\begin{figure}[t]
    \centerline{\includegraphics[width=0.2\textwidth]{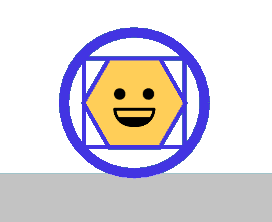}}
    \caption{Collision types preview. The thickest line indicates which collider is active and will react to the collision.}
    \Description{Collision types preview. The thickest line indicates which collider is active and will react to the collision.}
    \label{fig:collision}
\end{figure}

\begin{itemize}
    \item \textbf{SpriteRenderer2D} - Change appearance \\
    This component allows the player to select the appearance of the "Player"-object within the game.
    
    \item \textbf{Rigidbody2D} - Add physics \\ 
    The settings for this component can vary within a level to avoid overwhelming the player. For instance, the "Player"-object has mass and gravity scale settings, while the "Platform" object can be frozen in specific directions.

    \item \textbf{Collider2D} - Add collision detection \\
    The game offers three collider components: CircleCollider2D, BoxCollider2D, and PolygonCollider2D. Activating the \textit{show} option allows players to preview collisions, aiding in understanding object interactions. Enabling/ Disabling the component also changes the appearance of the preview and influences the line thickness. An example is illustrated in Figure~\ref{fig:collision}.

    \item \textbf{Scripts} - Add behavior \\
    To ensure accessibility, various pre-defined scripts are provided by the game, accessible via an icon next to the added component name. These scripts come with adjustable settings, such as \textit{speed} in the movement script and \textit{height} in the jump script. The settings may also interact with each other.

\end{itemize}

\subsection{Help View}

\begin{figure}[t]
    \centerline{\includegraphics[width=0.4\textwidth]{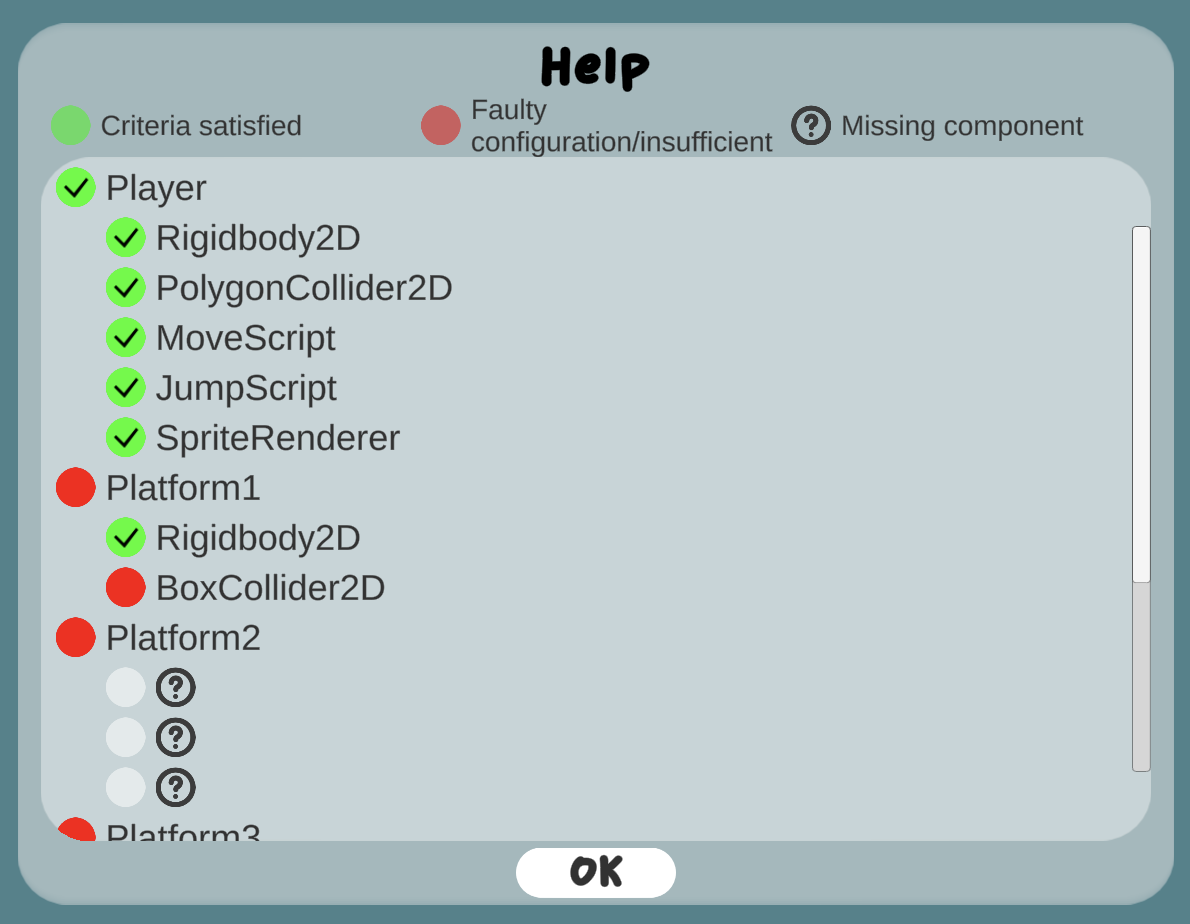}}
    \caption{The Help-View. The graphical representation provides a visual guide with additional information about missing or improperly configured components in the game.}
    \Description{The Help-View. The graphical representation provides a visual guide with additional information about missing or improperly configured components in the game.}
    \label{fig:help_view}
\end{figure}

\begin{figure}[t]
    \centerline{\includegraphics[width=0.4\textwidth]{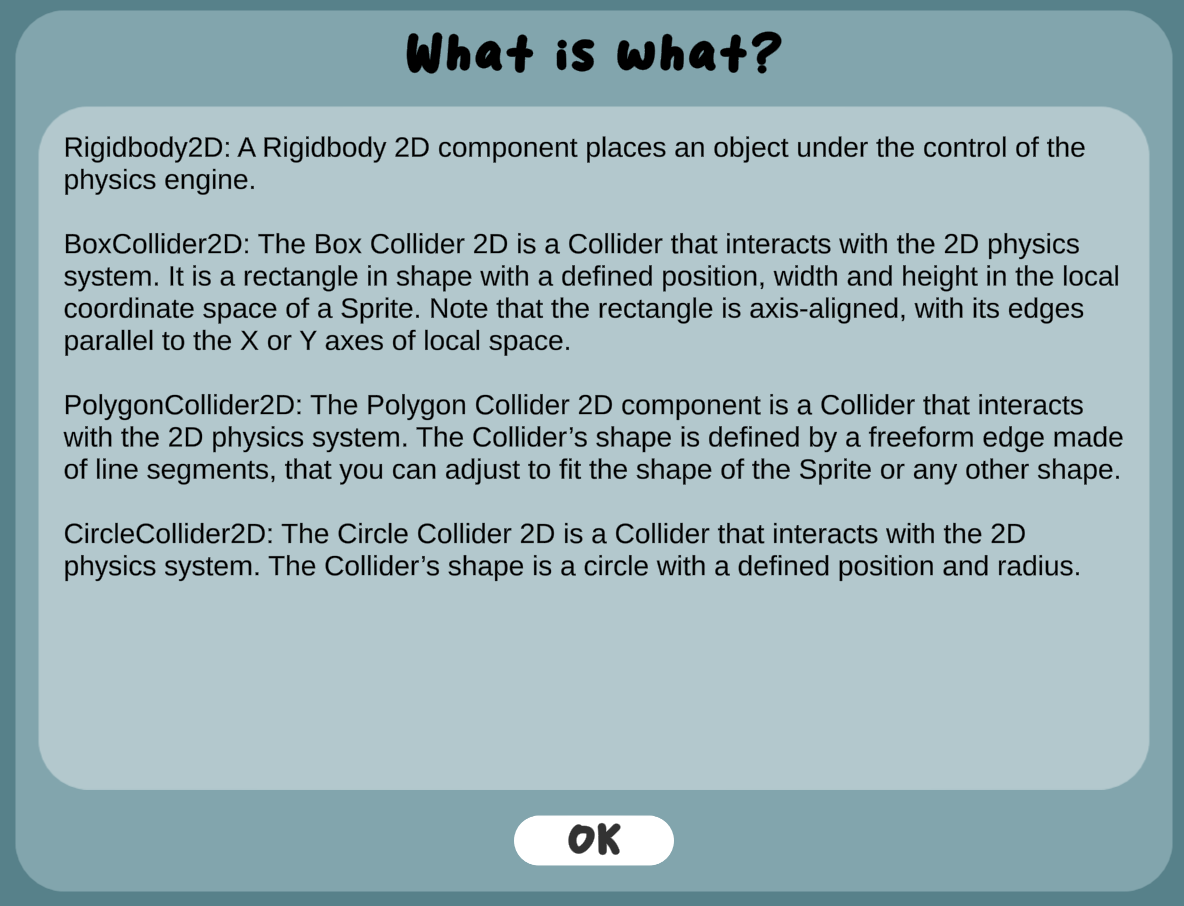}}
    \caption{The What-is-What-View. The popup provides detailed information about the available components for the selected game object.}
    \Description{The What-is-What-View. The popup provides detailed information about the available components for the selected game object.}
    \label{fig:wiw_view}
\end{figure}

When clicking on the "help"-button in the lower left corner, a popup presents additional information about missing components or faulty component configurations. Figure \ref{fig:help_view} shows an exemplary representation of the help view where each game object is listed with its accomplishment status and its required components. While a green marker indicates that all necessary components are added and correctly configured, a red indicator represents a game object that is not completely configured. For components, a green dot indicates accurate settings, while a red one signals that the component is correctly added but with a faulty configuration. Components showing an unknown status mean that the correct component has not yet been added. Incorrectly added components are shown to the user as a red indicator of the game object. 
%1.3
Additionally, a detailed description of each available component is accessible via the (?)-button next to the component drop-down. The respective popup is depicted in Figure \ref{fig:wiw_view}.

\section{Evaluation}

The study aimed to evaluate the efficacy of a game-based learning tool to teach concepts of game development. The evaluation was conducted as an AB user study to compare the game-based learning approach (Group A) with traditional learning methods (Group B). While participants in Group A started directly with the game, participants in Group B received a theoretical description in a written form. The research focus was on (1) user experience, (2) learning, (3) design, and (4) engagement. 

\subsection{Material and Setup}

\begin{figure}[t]
    \centerline{\includegraphics[width=0.5\textwidth]{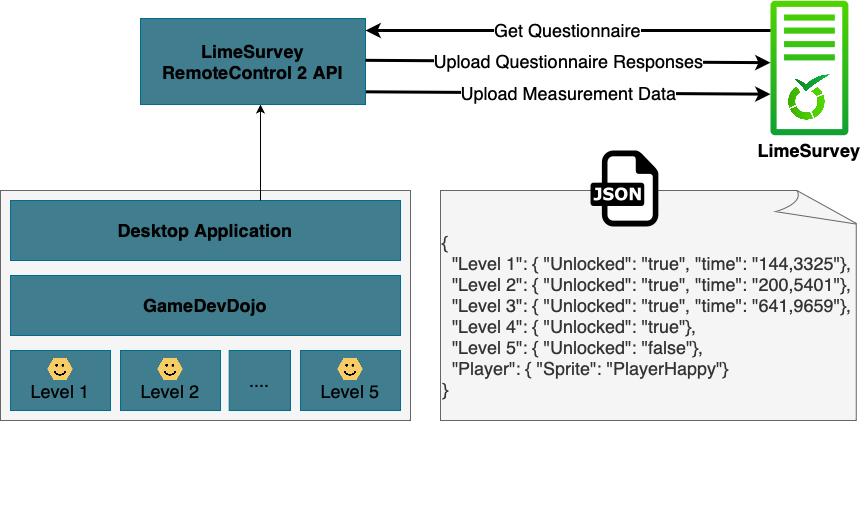}}
    \caption{Loading and uploading measurement data to LimeSurvey. To communicate with the server, the LimeSurvey RemoteControl 2 API is used.}
    \Description{Loading and uploading measurement data to LimeSurvey. To communicate with the server, the LimeSurvey RemoteControl 2 API is used.}
    \label{fig:limesurvey}
\end{figure}

To participate in the evaluation process, each participant was required to use a private computer running either Windows or Mac OS, or an Android tablet. Participants were asked to download and install the executable on their devices via an online cloud link. For the installation of the app, no specific hardware requirements were required. Apart from the operating system, the only prerequisite for running the application was a functional internet connection for the LimeSurvey integration. For loading and uploading the survey data, we used the LimeSurvey RemoteControl 2 API \cite{LimeSurvey2022}. Figure \ref{fig:limesurvey} shows the conceptual overview of the server communication for loading and uploading the survey data.

\subsection{Method and Procedure}
% 2.2
% 2.3
To conduct the study, we recruited 57 participants from two local schools. Each test run was designed as a 100-minute session in a classroom setting where students were randomly assigned to a test group. After assigning them to one group, they were asked to fill out a pre-questionnaire about their age, gender, experience in gaming and game development, as well as preferred gaming platforms. To measure the learning outcome, we asked participants ten knowledge questions before starting the learning activity and after completing the tasks. The knowledge questions included three single-choice and seven multiple-choice questions about general game development concepts. After filling out the pre-questionnaire and answering the knowledge questions, participants in Group A started playing the game. The gameplay included five different levels of game components like a rigid body, colliders, platforms, player movement, or collectibles. Participants in Group B received an 8-page PDF document with the same theoretical content and visual illustrations about game development concepts such as rigid body, collider, collision, and platform effector components. During the game, participants from Group A were asked to complete the following tasks in the levels:
\begin{enumerate}
    \item Add a gravity force to the GameObject and change the appearance.
    \item Collide the objects in the most precise way.
    \item Again, collide the objects in the most precise way but keep in mind the used memory.
    \item Make the player move and jump onto/through the platforms.
    \item Add a script to collect a coin.
\end{enumerate}
While performing the tasks in the game, we measured the learning time for each level as well as the total gameplay time. The time measurements were integrated into the learning environment and started automatically when the user loaded a level. After finishing the game or the theoretical text descriptions, both groups were asked the same knowledge questions again to assess their acquired knowledge during the learning sessions and to rate their interest in game development on a Likert scale between 1 (low interest) and 5 (high interest). Once the participants in Group B completed the post-knowledge questions, Group B was allowed to play the game as well.
While Group A was asked to play a certain amount of time, Group B was limited in time for playing the game. Due to time constraints, only a subset of Group B tried the game. Participants who played the game had a five-minute break after the second knowledge questionnaire.
In the end, participants were asked a post-questionnaire that included two standardized questionnaires to measure the user and learning experience. For evaluating the user experience, we used the Player Experience Inventory (PXI) questionnaire \cite{PXI_EN} where participants rated their experience on a Likert scale between -3 (strongly disagree) and 3 (strongly agree). To assess the user's impressions regarding their learning experience, the Web-Based Learning Tools (WBLT) questionnaire \cite{WBLT} was used. Participants were asked to rate their perceptions of the three constructs—learning, design, and engagement on a Likert scale from 1 (strongly disagree) to 5 (strongly agree). 

\subsection{Participants}

In total, 57 middle and high school students (45 females, 8 males, and 4 others) aged between 12 and 18 (AVG=14.6, SD=1.73) participated in the study and were divided into two groups:
25 students in Group A and 32 in Group B. The different group sizes resulted from the different class sizes in the schools and the willingness of the pupils to take part in the study.
31 participants had fundamental computer skills, 16 were familiar with gaming, and none had specific knowledge of game development.

\section{Results}

This section outlines the results derived from the 57 data sets gathered in the study. The findings are categorized into learning outcomes, game measurements, learning experience, and engagement. 

\subsection{Learning Outcome and Interest}

To assess the learning outcomes, participants were asked the same knowledge questions before and after the learning experiment. Among the participants in Group A, 28.0\% of the ten questions were correctly answered, while the participants in Group B achieved a success rate of 26.9\%. In comparison with the outcomes from the post-questionnaire, participants from Group A increased the number of accurately answered questions by 16.4\%, whereas Group B showed an improvement of 25.3\% in their knowledge outcomes. To detect significant changes within the paired knowledge data, we used the McNemar's test. Both groups significantly increased their knowledge outcomes. While Group A showed a significant improvement in three questions, participants in Group B performed significantly better on five questions after the learning experience. Table \ref{tab:question10} shows the success rate of the different groups, including the corresponding p-values from the significant test as well as the differences between the learning curves, measured by the increase-values between both groups. According to these values, it can be observed that Group B had a higher learning outcome on six questions, resulting in an overall higher learning curve. For testing the differences in each question for significance we used the Fishers Exact Test, which showed only one significant difference (Q8) in the behavior of Group B. To determine the overall significant difference in the learning increase between the two groups, we used the Wilcoxon Rank Sum Test (WRST) which showed no significant difference (p-value: 0.52).

Comparing the level of interest in the game development before and after the learning session showed that participants who used the learning application experienced a slight increase in interest (pre: AVG=2.60, SD=1.44; post: AVG=3.00, SD=1.46), while Group B showed a decrease in interest (pre: AVG=2.68, SD=1.49; post: AVG=2.34, SD=1.45). Accordingly, a reduced number of participants were able to image themselves doing game development as a hobby after performing in the theoretical learning group (pre: AVG=2.48, SD=1.53; post: AVG=2.19, SD=1.28). In contrast, participants who used the learning app showed a higher interest in game development as a leisure activity (pre: AVG=2.48, SD=1.45; post: AVG=2.56, SD=1.50).

\begin{table*}[t]
\centering
\caption{Grouped knowledge success rate and learning increase.}
\label{tab:question10}
\begin{tabular}{ccccccccccc}
\toprule
\multirow{2}{*}{\textbf{Question}} & \multicolumn{4}{c}{\textbf{Group A}} & \multicolumn{4}{c}{\textbf{Group B}} & \multicolumn{2}{c}{\textbf{Learning}} \\
\cmidrule(lr){2-5} \cmidrule(lr){6-9} \cmidrule(lr){10-11}
& \textbf{Pre} & \textbf{Post} & \textbf{Increase} & \textbf{P$_{McNemar}$} & \textbf{Pre} & \textbf{Post} & \textbf{Increase} & \textbf{P$_{McNemar}$} & \textbf{Diff} & \textbf{P$_{Fisher}$} \\ \hline
Q1 & 0.0\% & 8.0\% & 8.0\% & 0.480 & 3.1\% & 9.4\% & 6.2\% & 0.617 &\textbf{1.8\%} & 1\\
Q2 & 24.0\% & 76.0\% & 52.0\% & 0.002 $\ast$ & 12.5\% & 59.4\% & 46.9\% & 0.001 $\ast$ &\textbf{5.1\%} & 0.792\\
Q3 & 0.0\% & 0.0\% & 0.0\% & - & 0.0\% & 3.1\% & 3.1\% & 1 &-3.1\% & 1\\
Q4 & 20.0\% & 48.0\% & 28.0\% & 0.070 & 15.6\% & 46.9\% & 31.2\% & 0.009 $\ast$ &-3.2\% & 1 \\
Q5 & 0.0\% & 4.0\% & 4.0\% & 1 & 0.0\% & 12.5\% & 12.5\% & 0.134 &-8.5\% & 0.372\\
Q6 & 4.0\% & 28.0\% & 24.0\% & 0.041 $\ast$ & 3.1\% & 40.6\% & 37.5\% & 0.003 $\ast$ &-13.5\% & 0.391\\
Q7 & 12.0\% & 16.0\% & 4.0\% & 1 & 6.2\% & 15.6\% & 9.4\% & 0.371 &-5.4\% & 0.623\\
Q8 & 8.0\% & 16.0\% & 8.0\% & 0.048 & 3.1\% & 40.6\% & 37.5\% & 0.001 $\ast$ &-29.5\% & 0.013 $\ast$\\
Q9 & 4.0\% & 4.0\% & 0.0\% & - & 3.1\% & 0.0\% & -3.1\% & 1 & \textbf{3.1\%} & -\\
Q10 & 12.0\% & 44.0\% & 32.0\% & 0.027 $\ast$ & 9.4\% & 31.2\% & 21.9\% & 0.046 $\ast$ &\textbf{10.1\%} & 0.546\\
\cmidrule{1-11}
Total & 28.0\% & 44.4\% & 16.4\% & 3.479e$^{-08}$ $\ast$ & 26.9\% & 52.2\% & 25.3\% & 5.996e$^{-13}$ $\ast$ & -8.9\% & 0.52 $_{WRST}$\\
\midrule
\multicolumn{11}{p{15cm}}{Q1: A GameObject is ...}\\
\multicolumn{11}{p{15cm}}{Q2: To simulate the effect of gravity on a GameObject one has to use the following component ...}\\
\multicolumn{11}{p{15cm}}{Q3: Which of the following properties of GameObjects can be manipulated using Rigidbody and Transform components?}\\
\multicolumn{11}{p{15cm}}{Q4: Which type of collider is used to circumference complexly shaped GameObjects in a precise manner?}\\
\multicolumn{11}{p{15cm}}{Q5: When does a collision occur?}\\
\multicolumn{11}{p{15cm}}{Q6: To keep memory usage at a minimum when colliding objects, one typically uses which of the following colliders?}\\
\multicolumn{11}{p{15cm}}{Q7: What are scripts?}\\
\multicolumn{11}{p{15cm}}{Q8: What are Effector components?}\\
\multicolumn{11}{p{15cm}}{Q9: Scripts are used for more complex behavior: Which of the following concepts can be used to collect GameObjects with another GameObject?}\\
\multicolumn{11}{p{15cm}}{Q10: Which of the following components defines a GameObjects visual appearance?}\\
\bottomrule
\end{tabular}
\end{table*}

\subsection{Game Time-Measurements}

\begin{figure}[t]
    \centering
    \includegraphics[width=0.5\textwidth]{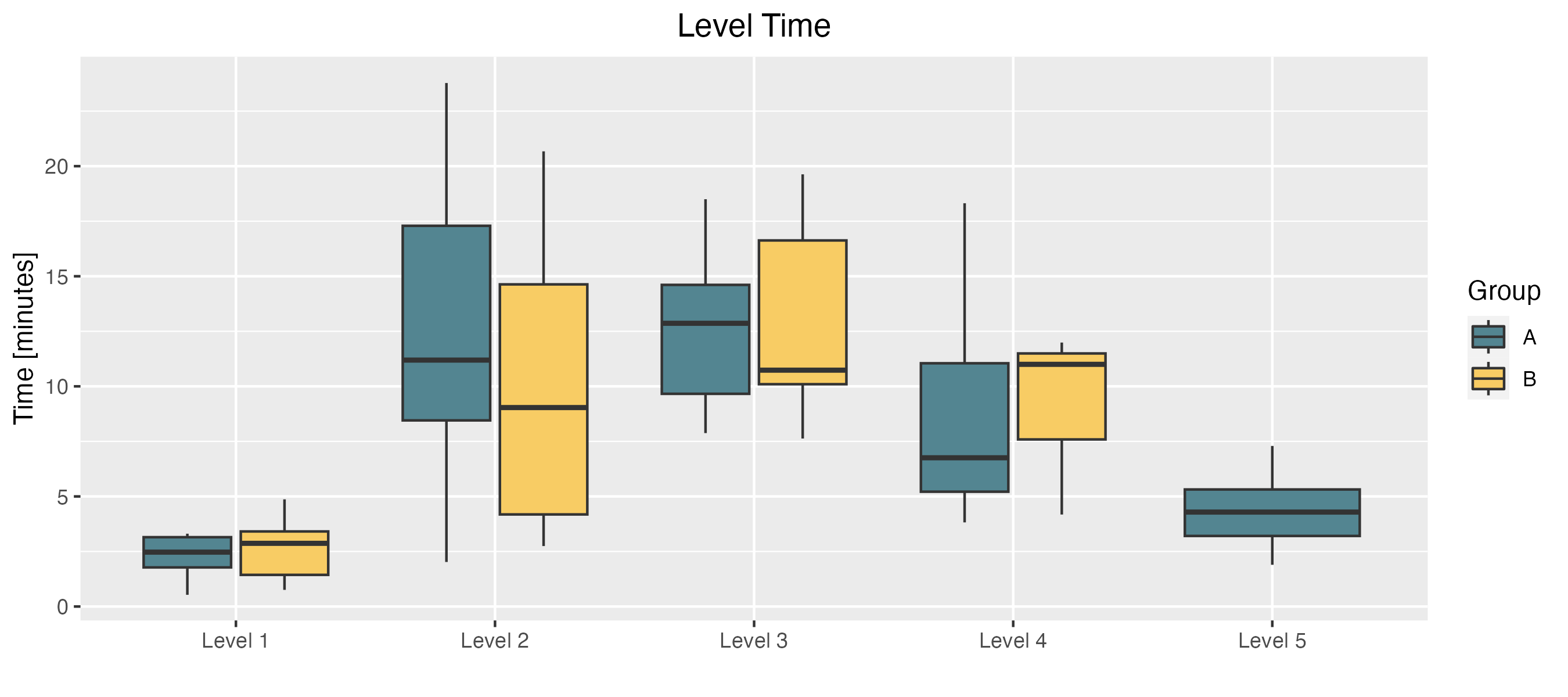}
    \caption{Compared average time per level.}
    \Description{Compared average time per level.}
    \label{fig:leveltime}
\end{figure}

While playing the game, we recorded the participant's level completion time. The outcomes of the time measurements, categorized by the two groups, are shown in Figure \ref{fig:leveltime}. While the quantile margin time for Group B was consistently shorter throughout the different game levels, the average completion time was higher in the more advanced levels. Even though Group B had a theoretical background at the time of playing the game, there were no significant differences detected via the Wilcoxon rank-sum test between the two groups. Table \ref{tab:levelstat} gives an overview of level completion times and the success rate by the groups. While more participants with prior theoretical knowledge completed the first two levels, they got outperformed in the more advanced levels. Especially for the 5th level, none of the participants from Group B were able to complete the level.

\begin{table}[t]
\centering
\caption{Grouped success rates over played levels ($B^* \subset B)$.}
\label{tab:levelstat}
\begin{tabular}{ccccc}
\toprule
\multicolumn{1}{c}{}&\multicolumn{1}{c}{}& \multicolumn{2}{c}{\textbf{Time}} & \multicolumn{1}{c}{\textbf{Players}}\\ %\cline{4-7}
\multicolumn{1}{l}{\multirow{-2}{*}{\textbf{Group}}} & \multicolumn{1}{c}{\multirow{-2}{*}{\textbf{Level}}} &\textbf{AVG} &\textbf{SD} & \multicolumn{1}{l}{\textbf{Completed}} \\ 
\midrule
                    & Level 1   & 3.16	  & 2.53    & 92.0\%  \\ %\cline{4-6} 
                    & Level 2   & 12.99	  & 7.73    & 88.0\%  \\ %\cline{4-6} 
                    & Level 3   & 14.52	  & 7.63    & 84.0\%  \\ %\cline{4-6} 
                    & Level 4   & 8.52	  & 4.54    & 60.0\%  \\ %\cline{4-6} 
\multirow{-5}{*}{A} & Level 5   & 4.55	  & 2.09    & 48.0\%  \\ \midrule
                    & Level 1   & 2.65    & 1.30    & 93.3\%  \\ %\cline{4-6} 
                    & Level 2   & 10.05   & 6.63    & 93.3\%  \\ %\cline{4-6} 
                    & Level 3   & 17.67   & 15.56   & 46.7\%  \\ %\cline{4-6} 
                    & Level 4   & 9.06    & 4.25    & 20.0\%  \\ %\cline{4-6} 
\multirow{-5}{*}{B*} & Level 5   & -       & -       & 0.0\%   \\ 

\bottomrule
\end{tabular}
\end{table}

\subsection{Learning Experience and Motivation}

To gain further insight into users' perceptions of the learning application, participants were asked to rate their experience on a 7-point Likert scale from -3 (strongly disagree) to 3 (strongly agree). The assessment consists of ten different constructs, each consisting of three interrelated questions. Table \ref{tab:pxi} shows the evaluation outcomes of the different constructs derived from the PXI questionnaire. The results indicate a moderately high level for almost all categories. Among the constructs, the audiovisual appeal was scored as the highest, followed by goal clarity, progress feedback, and immersion. In total, the participants mostly agreed with the statement "I was fully focused on the game." (AVG=2.25, SD=2.11). The categories of mastery and meaning obtained lower average scores, suggesting that there is still potential for improvement in terms of users' sense of skill development and the perceived significance of their interactions within the learning application.

To assess the learning experience and the effectiveness of the learning application design, we used the WBLT questionnaire. Participants were asked to rate the categories learning, engagement, and design on a Likert scale from 1 (strongly disagree) and 5 (strongly agree). The high learning score (AVG=4.23, SD=1.28) reflects participants' perceptions of the effectiveness of the learning application in improving understanding and knowledge acquisition. Furthermore, participants found the learning application engaging and would like to use the learning object again (AVG=4.38, SD=1.43). Regarding the design, participants rated the features in the learning object as useful and easy to use (AVG=4.52, SD=1.16).

\begin{table}[t]
\centering
\caption{Results of the Player Experience Inventory on a Likert Scale between -3 (strongly disagree) and 3 (strongly agree).}
\label{tab:pxi}
\begin{tabular}{lccc}
\hline
\textbf{Question Category} & \textbf{AVG} & \textbf{SD} \\%& \textbf{SUM} \\ 
\hline
Meaning                    & 1.11	      & 1.92	    \\%& 2.67         \\
Curiosity                  & 1.39	      & 2.02	    \\%& 9.33         \\
Mastery                    & 1.01	      & 2.16	    \\%& 0.33         \\
Autonomy                   & 1.19	      & 1.93	    \\%& 4.67         \\
Immersion                  & 1.61	      & 1.81	    \\%& 14.67        \\
Progress Feedback          & 1.63	      & 1.71	    \\%& 15.00        \\
Audiovisual Appeal         & 1.86	      & 2.01	    \\%& 20.67        \\
Challenge                  & 1.60	      & 1.85	    \\%& 14.33        \\
Ease Of Control            & 1.19	      & 2.16	    \\%& 4.67         \\
Clarity Of Goals           & 1.69	      & 1.95	    \\%& 16.67        \\
Enjoyment                  & 1.39	      & 2.11	    \\%& 9.33         \\
\hline 
\end{tabular}
\end{table}

%\begin{table}[!ht]
%\centering
%\begin{tabular}{lccc}
%\hline
%\textbf{Question} & \textbf{AVG} & \textbf{SD} \\%& \textbf{SUM}   \\ 
%\hline
%Learning          & 4.23        & 1.28        \\%& 5.4            \\
%Design            & 4.52        & 1.16        \\%& 12.5           \\
%Engagement        & 4.38        & 1.43        \\%& 9.0            \\ 
%\hline      
%\end{tabular}
%\caption{Results of the WBLT questionnaire on a Likert Scale between 1 (strongly disagree) and 5 (strongly agree)}
%\label{tab:wblt}
%\end{table}

\subsection{General Comments}

Participants' comments reflect a generally positive sentiment towards the game, with many expressing enjoyment, excitement, and amusement during gameplay. Some found the game fun, cool, and entertaining, even for those unfamiliar with such activities. However, not all participants found the concept equally engaging, as some felt it did not resonate with their interests or did not find it particularly entertaining. Several participants appreciated the learning opportunity the game provided, especially for newcomers to programming and game development. Others highlighted the educational value, describing it as a good way to gain insight into game development by changing the game components individually. Overall, participants expressed a mix of enjoyment, learning, and varying degrees of interest in the game development experience.

\section{Discussion}

The study aimed to explore the learning outcome as well as the learning experience and motivation of a game-based learning approach to teach fundamental concepts of game development. The evaluated learning outcomes showed a remarkable improvement in correctly answered knowledge questions after performing the learning application. Although the participants in one group had theoretical knowledge, no statistically significant differences in completion times were found between the participants with and without prior knowledge. While participants with theoretical knowledge performed better at initial levels, their performance lagged at more advanced levels. 
%3.1
Considering the time cut-off for Group B, their performance may have been impacted by factors such as time pressure, stress, or a rushed decision-making process. This time constraint could have hindered their ability to apply their theoretical knowledge effectively.
An additional factor could be that users get bored when reading long text descriptions, in comparison to users playing a game \cite{prensky2003digital}. 
Nevertheless, participants with the purely textual description performed better in the knowledge questions and had a higher increase in their knowledge. However, Bloom's taxonomy suggests that lower cognitive stages like remembering and recalling facts represent only fundamental cognitive abilities, whereas higher cognitive stages like applying and using the learned content in novel situations involve more complex thought processes \cite{bloom2020taxonomy}.
Participants who were actively involved with the learning application were able to solve the more advanced tasks faster, which could indicate that they reached a higher cognitive level than the participants who were previously provided with theoretical knowledge. The PXI score for audiovisual appeal indicates that users generally found visual and auditory elements appealing and engaging. Furthermore, the clarity of goals was also well received, suggesting that participants found the learning objectives and goals of the application easily understandable. In contrast, the moderate score for mastery, even though it still retained a positive range, implies that users may have felt relatively less confident in their perceived mastery of the application's content. Furthermore, the results highlight that users expressed satisfaction with the progress feedback aspect, suggesting that they received meaningful and informative feedback on their progress within the application. It further indicates that the confetti rain feedback and level of achievements had a positive effect. This is consistent with the findings of Elliot et al. \cite{elliot2005conceptual} who highlighted in their work that achievement-oriented goals have a significant effect on the motivation in accomplishing tasks. Moreover, Gurbuz et al. \cite{gurbuz2022serious} already pointed out, that immersive elements are crucial factors for motivating and engaging players throughout the game. The results on the immersion score of the PXI are in line with that and show that users were deeply engaged in the learning content and activities. This underscores that incorporating achievement-oriented mechanisms and immersive elements in serious games can contribute to increased motivation, engagement, and more effective learning outcomes, especially for game development.
These findings also correlate with the high engagement score from the WBLT questionnaire, which shows consistent results. Additionally, the design construct with the highest score from the WBLT shows congruent user impressions for progress feedback and audiovisual appeal.
Comber et al. \cite{comber2019engaging} have shown that game engines are not an effective approach to teaching low-level instructions for game development. The study results demonstrated that a game-based learning approach for teaching game development concepts can be an efficient way to engage people in developing games. The presented concept could be an additional improvement in acquiring fundamental knowledge before teaching how to develop games with a professional game engine. 

\subsection{Limitations}

While this study offers valuable insights into a user's learning experience and motivation for a game-based learning tool for teaching game development concepts, some limitations have to be considered. One of the limitations of this study was the relatively small sample size of 57 participants. The number of participants involved in the study might not fully reflect the diverse range of K-12 students. Furthermore, the reliance on self-report measures introduces the possibility of self-report bias. Participants may not always provide entirely accurate or honest responses due to social desirability or recall limitations. This bias could influence the reported perceptions. The study was conducted in a classroom setting and was limited to a time frame of 100 minutes. Longer-term effects, fluctuations, or changes that could manifest beyond the study period might not have been captured due to this constraint.
\section{Conclusion}

This study investigated the impact of an educational game approach on learning outcomes, motivation, and overall user experience in teaching game development concepts. The results show a potential way to link a game-based learning approach with game development concepts. The assessed learning outcomes demonstrated a significant improvement in correctly answered knowledge questions after performing the learning application. While participants who were provided with theoretical knowledge had a better performance in the knowledge questions, it was seen that participants who were engaged with the learning application were able to solve the more advanced tasks faster. Moreover, the learning application had a positive influence on participants' interest in game development. This suggests that the interactive and hands-on nature of the learning application contributed to increasing participants' enthusiasm for game development as a leisure activity. The general comments from the participants indicated a diverse range of sentiments, from enjoyment and enthusiasm to appreciation for the learning opportunity provided by the learning application. Overall, the study underscores the effectiveness of game-based learning in enhancing learning outcomes and engagement in the field of game development education. The presented results contribute to an evolving understanding of how interactive learning methodologies, particularly in the context of game development, can shape more effective and engaging educational experiences. In addition, future work could focus on more gamification elements to increase the validity and generalizability of the results. The integration of gamification elements, like collecting stars based on completion time or number of help requests, could provide valuable insights into the motivational aspects of learning. The introduction of badges or achievements for completing components without requesting help could also be explored. Furthermore, examining the long-term effects of these gamified elements and comparing different learning methods could provide a comprehensive understanding of their impact on performance and retention over time.
\pagebreak

%%
%% The next two lines define the bibliography style to be used, and
%% the bibliography file.
\bibliographystyle{ACM-Reference-Format}
\bibliography{references}

\end{document}